\documentclass[conference]{IEEEtran}
\IEEEoverridecommandlockouts
\usepackage{cite}
\usepackage{amsmath,amssymb,amsfonts}
\usepackage{algorithmic}
\usepackage{graphicx}
\usepackage{textcomp}
\usepackage{xcolor}

\usepackage{subfigure}
\usepackage[ruled,linesnumbered, vlined]{algorithm2e}
\usepackage{enumitem}
\usepackage{multirow} 
\usepackage{diagbox}
\usepackage[colorlinks=true, linkcolor=blue, citecolor=blue, urlcolor=blue]{hyperref}

\def\BibTeX{{\rm B\kern-.05em{\sc i\kern-.025em b}\kern-.08em
    T\kern-.1667em\lower.7ex\hbox{E}\kern-.125emX}}
\begin{document}

\title{Ghidorah: Fast LLM Inference on Edge with Speculative Decoding and Hetero-Core Parallelism
}
\author{
Jinhui Wei\textsuperscript{1}, Ye Huang\textsuperscript{1}, Yuhui Zhou\textsuperscript{1}, Jiazhi Jiang\textsuperscript{2}, Jiangsu Du\textsuperscript{1*}, Yutong Lu\textsuperscript{1} \\
\textsuperscript{1}Sun Yat-sen University, China, \textsuperscript{2}Beijing Normal University, China  \\
% \textsuperscript{*}Co-first authors, \textsuperscript{§}Corresponding author
}

\maketitle
\begingroup

\renewcommand\thefootnote{}\footnotetext{
Jinhui Wei and Ye Huang contributed equally to this work and are considered co-first authors.
Yuhui Zhou participated in this work as a research intern at Sun Yat-sen University.
* Corresponding author: Jiangsu Du (dujiangsu@mail.sysu.edu.cn).
}
\endgroup

\begin{abstract}
In-situ LLM inference on end-user devices has gained significant interest due to its privacy benefits and reduced dependency on external infrastructure.
However, as the decoding process is memory-bandwidth-bound, the diverse processing units in modern end-user devices cannot be fully exploited, resulting in slow LLM inference.
This paper presents Ghidorah, a LLM inference system for end-user devices with the unified memory architecture.
The key idea of Ghidorah can be summarized in two steps: 
1) leveraging speculative decoding approaches to enhance parallelism, and
2) ingeniously distributing workloads across multiple heterogeneous processing units to maximize computing power utilization.
Ghidorah includes the hetero-core model parallelism (HCMP) architecture and the architecture-aware profiling (ARCA) approach.
The HCMP architecture guides partitioning by leveraging the unified memory design of end-user devices and adapting to the hybrid computational demands of speculative decoding.
The ARCA approach is used to determine the optimal speculative strategy and partitioning strategy, balancing acceptance rate with parallel capability to maximize the speedup.
Additionally, we optimize sparse computation on ARM CPUs.
Experimental results show that Ghidorah can achieve up to 7.6$\times$ speedup in the dominant LLM decoding phase compared to the sequential decoding approach in NVIDIA Jetson NX.
\end{abstract}

\begin{IEEEkeywords}
Edge Intelligence, LLM Inference, Speculative Decoding, Unified Memory Architecture
\end{IEEEkeywords}

\section{Introduction}

Deploying large language models (LLMs) directly on end-user devices, such as mobile phones and laptops, is highly attractive.
In-situ inference can eliminate data privacy concerns and reduce reliance on external infrastructures, thereby boosting intelligence and autonomy.
However, achieving fast LLM inference remains a significant challenge, constrained by both algorithmic parallelism and hardware capability.

\begin{figure}[!t]
    \centering 
    \includegraphics[width=0.42\textwidth]{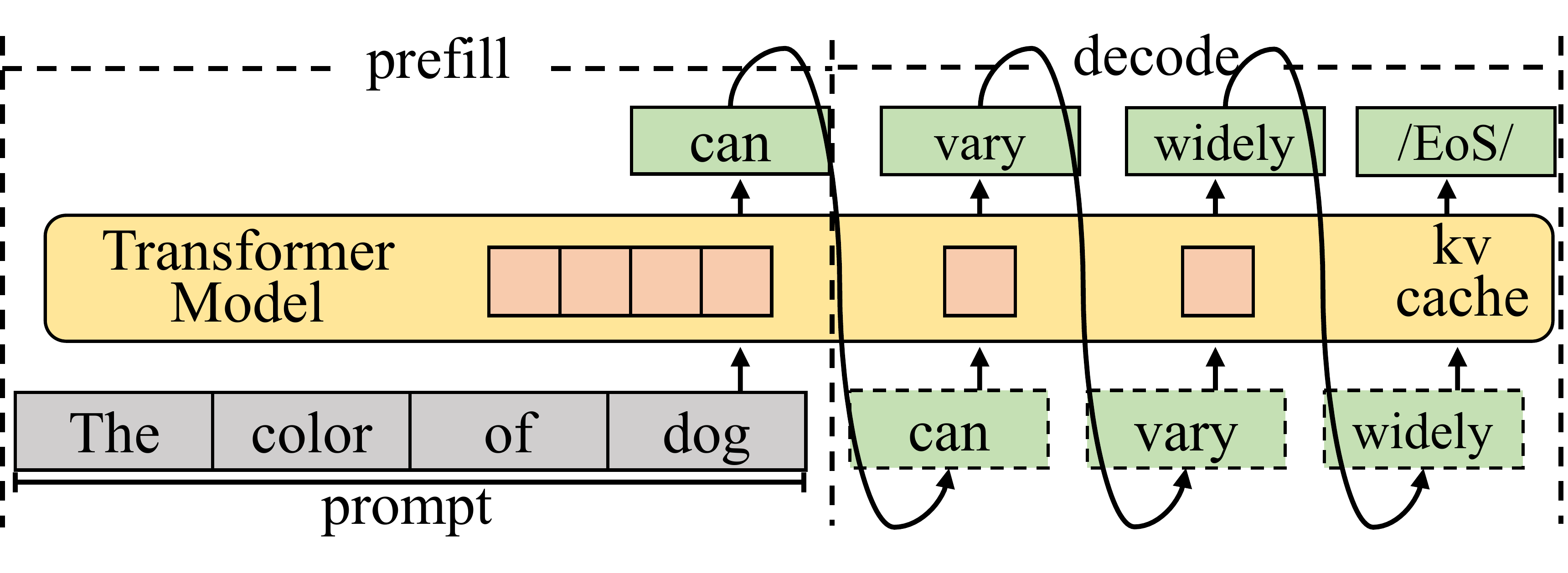}
    \vspace{-10pt}
    \caption{The autoregressive process of LLM inference with the KV cache technique. The decoding phase generally dominates the execution time.}
    \vspace{-5pt}
    \label{fig:gentask}
\end{figure}

To begin with, end-user devices typically process only a single request at a moment and the traditional decoding approach, shown in Figure~\ref{fig:gentask}, generates tokens sequentially, one at a time.
With the KV cache technique~\cite{kwon2023efficient}, the dominant decoding phase of the LLM inference has very limited parallelism, while requiring loading all the model weights in each iteration.
Thus, the limited parallelism and memory-bandwidth-bound nature prevent the full utilization of modern hardware.
Alternatively, the speculative decoding approach~\cite{specinfer, medusa, lookahead_decoding} takes a draft-then-verify decoding process.
In each decoding step, speculative decoding first drafts multiple tokens as predictions for future steps and then verifies in parallel whether to accept the drafted tokens.
It can parallelize the inherently sequential process, showing strong potential to accelerate token generation rates by trading off compute for bandwidth.

After enhancing the algorithmic parallelism, the LLM inference starts to be blocked by the limited hardware capability of end user devices.
To fully leverage the capabilities of end-user devices, it is promising to distribute workloads across heterogeneous processing units~\cite{codl, zhang2023edgenn, wang2019highhigh, Dopia2022}.
For instance, the GPU of Apple M4~\cite{appleM4} Pro has 17.04 TFLOPs FP16 performance, while its CPU has around 38 TFlops FP16 performance.
However, existing systems~\cite{vllm, medusa} do not align with the computing pattern of speculative decoding and the unified memory architecture of end-user devices, resulting in inefficiency.

First, existing model partitioning approaches~\cite{shoeybi2019megatron, ye2024galaxy} for distributing workloads across devices focus excessively on minimizing inter-device communication overhead and overlook opportunities for different processing units to access the same memory.
When distributing LLM inference across multiple devices, explicit communication is required, making the reduction of communication overhead a top priority.
In comparison, end-user devices typically integrate multiple heterogeneous processing units on a single chip and organize them within a unified memory architecture.
Well-known examples include Apple’s M-series SoCs~\cite{appleM4} and Intel Core Ultra~\cite{intelcoreultra}, and NVIDIA Jetson~\cite{nvjetson}.
In this way, minimizing memory access by carefully organizing partitioning strategy and data placement becomes the top priority for the unified memory architecture.

% ybrid computational demands of specu-
% lative decoding

Second, existing systems treat speculative decoding the same as the traditional decoding with a large batch size, less considering its specific computing pattern, particularly hybrid computational demands.
The attention mechanism in the traditional decoding calculates the correlation between every pair of tokens, while it only requires to calculate correlation between partial pair of tokens in speculative decoding.
Given the extremely high capability of accelerators in cloud systems and the high cost of transferring data to other processing units in the discrete memory architecture, this sparsity is often handled as dense computation using a mask mechanism.
This oversight leads to missed opportunities to leverage computing affinity with processing units, such as on CPU.

Third, algorithmic studies on speculative decoding primarily focus on increasing the acceptance length (the average number of tokens accepted in a single step), overlooking the balance between acceptance length and limited computing resources.
Verifying more combinations of drafted tokens leads to better acceptance length, while resulting in more FLOPs to generate a single token.
Compared to cloud hardware, end-user devices offer fewer resources for parallel verification, and verifying too many combinations can lead to decreased performance.
Besides, modern hardware widely leverages vectorization technique, enabling the parallel processing of multiple data elements.
It is crucial to select an appropriate verification width tailored to the specific hardware, thereby achieving the optimal balance between acceptance length and hardware parallelism.

In this paper, we introduce Ghidorah, a system specifically designed to accelerate single-sample LLM inference in end-user devices with a unified memory architecture. 
It explores speculative decoding approaches to tackle the memory-bandwidth limitations inherent in the single-sample decoding process.
Next, it distributes the novel workloads across multiple heterogeneous processing units in end-user devices, taking into account the hardware architecture and computing patterns.
In summary, we make the following contributions:
\begin{itemize}
    \item We identify the performance bottlenecks in the single-sample LLM inference and explore speculative decoding approaches.
    \item We present Ghidorah, a large language model inference system for end-user devices. It incorporates the hetero-core model parallelism (HCMP) architecture and the architecture-aware profiling (ARCA) approach to better employ the speculative decoding and align with the unified memory architecture.
    \item We implement customized sparse computation optimizations for ARM CPUs.
    % \item We propose the HCMP architecture to guide the model partitioning across heterogeneous processing units. Benefiting from the zero-copy technique, it incurs minimal data synchronization overhead and allocates workloads to processing units with better affinity. It also implements customized sparse computation optimizations for ARM CPUs.
    % \item We propose the ARCA approach to determine the speculative strategy and partitioning strategy to achieve the optimal balance between acceptance length and hardware parallelism.
\end{itemize}
Experimental results show that Ghidorah achieves up to a 7.6$\times$ speedup over the sequential decoding approach on the Nvidia Jetson NX. This improvement is attributed to a 3.27$\times$ algorithmic enhancement and a 2.31$\times$ parallel speedup.

% \item We present Ghidorah, a large language model inference system for end-user devices. It explores speculative decoding approaches to enhance single-sample parallelism and optimizes performance by considering both the hardware architecture and computing patterns.   
%     \item We propose hetero-core model parallelism to improve the distribution of workloads across multiple heterogeneous processing units organized within a unified memory architecture. This approach enables more frequent synchronization through the use of a zero-copy technique. Additionally, it employs online softmax to parallelize workloads with better affinity across different processing units.
%     \item We summarize the sparsity present in mainstream speculative decoding approaches and develop a customized kernel for efficient processing the sparse matrix multiplication on CPUs.
%     \item We propose the dynamic partitioning approach to address load imbalance among heterogeneous processing units as the context length incrementally increases during LLM inference.
\section{Background and Motivation}

% \begin{figure}[!t]
%     \centering 
%     \includegraphics[width=0.3\textwidth]{figure/gentask.pdf}
%     \vspace{-10pt}
%     \caption{The inference process of LLM generative task.}
%     % \vspace{-10pt}
%     \label{fig:gentask}
% \end{figure}

\subsection{LLM Generative Task Inference}

LLMs are primarily used for generative tasks, where inference involves generating new text based on a given prompt.
As shown in Figure~\ref{fig:gentask}, it predicts the next token and performs iteratively until meeting the end identifier (EoS).
By maintaining the KV cache in memory, modern LLM serving systems eliminates most redundant computations during this iterative process. The inference process is divided into prefill and decode phases based on whether the KV cache has been generated.
The prefill phase handles the newly incoming prompt and initializes the KV cache.
Since the prompt typically consists of many tokens, it has high parallelism.
Next, with the KV cache, each step of the decode phase processes only one new token generated by the previous step and appends the new KV cache.
In end-user scenarios, single-sample inference processes only a single token during each iteration of the decoding phase.
The decoding phase is heavily constrained by the memory bandwidth and cannot fully exploit hardware capability.
Moreover, the decoding phase typically dominates the overall inference process.

\subsection{Collaborative Inference and Transformer Structure}
Distributing inference workload across various devices or processing units is a common strategy for enhancing performance.

\subsubsection{Data, Pipeline and Sequence Parallelism}
Data parallelism (DP)~\cite{li2014communication} partitions workloads along the sample dimension, allowing each device to perform inferences independently.
Pipeline parallelism (PP)~\cite{wang2019highhigh} horizontally partitions the model into consecutive stages along layer dimension, with each stage placed in a device.
Since DP and PP are mainly used in high-throughput scenarios and cannot reduce the execution time for a single sample, they are not suitable for end-user scenarios.
Sequence parallelism (SP)~\cite{jacobs2023deepspeed, ye2024galaxy} is designed for the long-context inference and it divides the input along the sequence dimension.
It requires each device to load the entire model weights, making it advantageous only for extremely long sequences.
We do not consider it within the scope of our work.

\subsubsection{Tensor Parallelism and Transformer Structure}
\begin{figure}[!t]
    \centering 
    \includegraphics[width=0.45\textwidth]{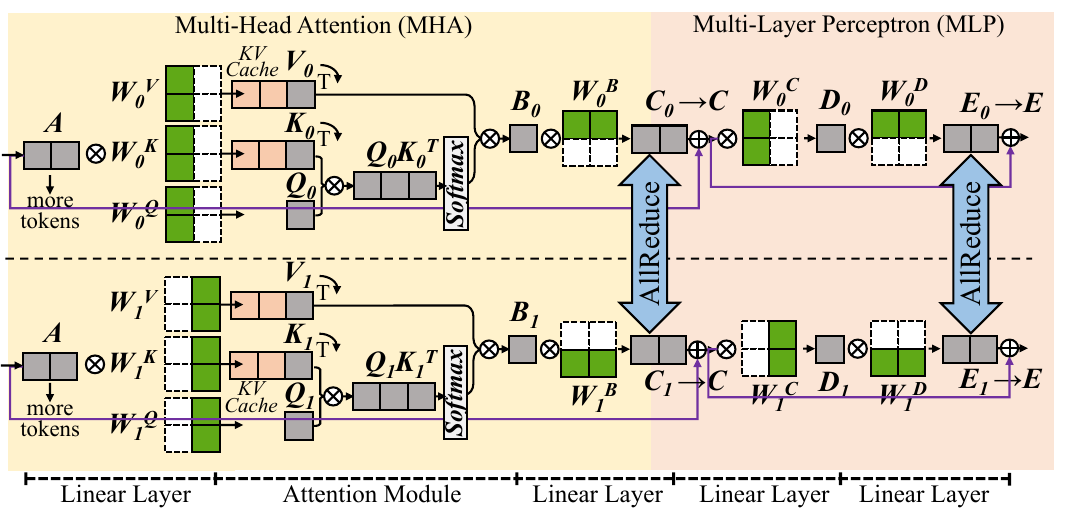}
    \vspace{-10pt}
    \caption{The partitioning~\cite{shoeybi2019megatron} of Transformer model in the matrix form.}
    \vspace{-5pt}
    \label{fig:llminfer}
\end{figure}

Tensor parallelism partitions model weights to different devices~\cite{shoeybi2019megatron, ye2024galaxy} or processing units~\cite{zhang2023edgenn, jia2022codl}, and it is promising to reduce the execution time of a single sample.
The Transformer~\cite{vaswani2017} structure is the primary backbone for LLMs, consisting of multiple stacked Transformer layers, all configured identically.
In a Transformer layer, the primary components are the multi-head attention (MHA) block and the multi-layer perceptron (MLP) block. 
Here we ignore element-wise operations, such as dropout, residual addition, and layernorm.
In MHA block, the first linear layer generates query (Q), key (K), and value(V) matrices for each attention head.
Each head conducts self-attention as in Equation~\ref{eq:qkv}, independently, and their outputs are concatenated and further processed by the next linear layer.
MLP block involves two consecutive linear layers.

\begin{equation}
\begin{aligned}
    & X = QK^T \\
    & A = softmax(X) \\
    & O = AV 
\end{aligned}
\label{eq:qkv}
\end{equation}

Figure~\ref{fig:llminfer} shows a single Transformer layer and it is partitioned using the cloud TP solution in Megatron-LM~\cite{shoeybi2019megatron}.
For every two consecutive linear layers, it partitions the weights of the first one by columns and those of the second one by rows.
An AllReduce communication is inserted as the synchronization point for every two linear layers.
For the attention module in MHA block, it is partitioned by heads, with different attention heads assigned to different devices.

\subsection{Speculative Decoding}

\begin{figure}[!t]
    \centering
    \subfigure[Medusa Verification Tree.]{
        \includegraphics[width=0.23\textwidth]{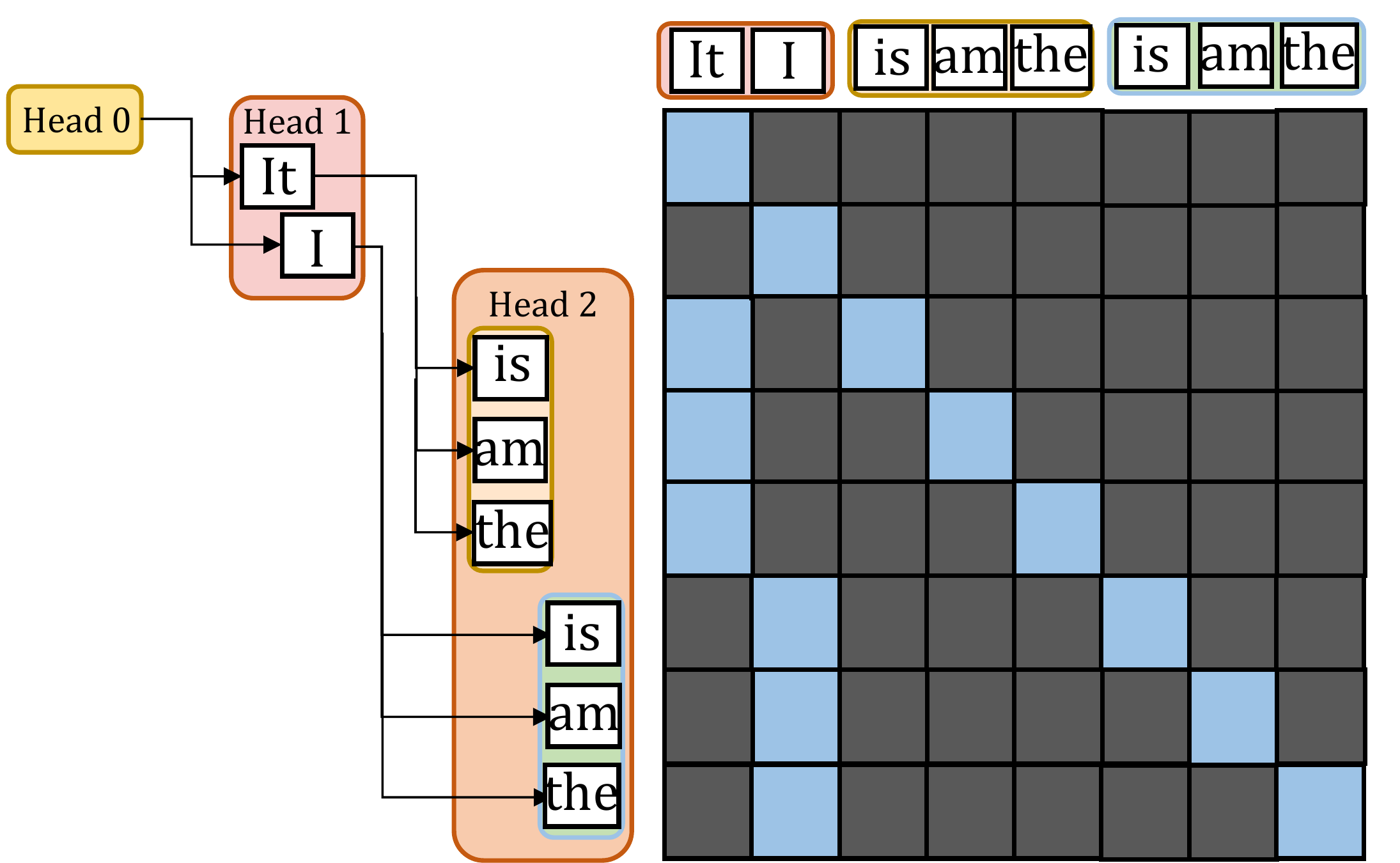}
        \label{fig:mask_tree}
    }
    \subfigure[Medusa~\cite{medusa}.]{
        \includegraphics[width=0.18\textwidth]{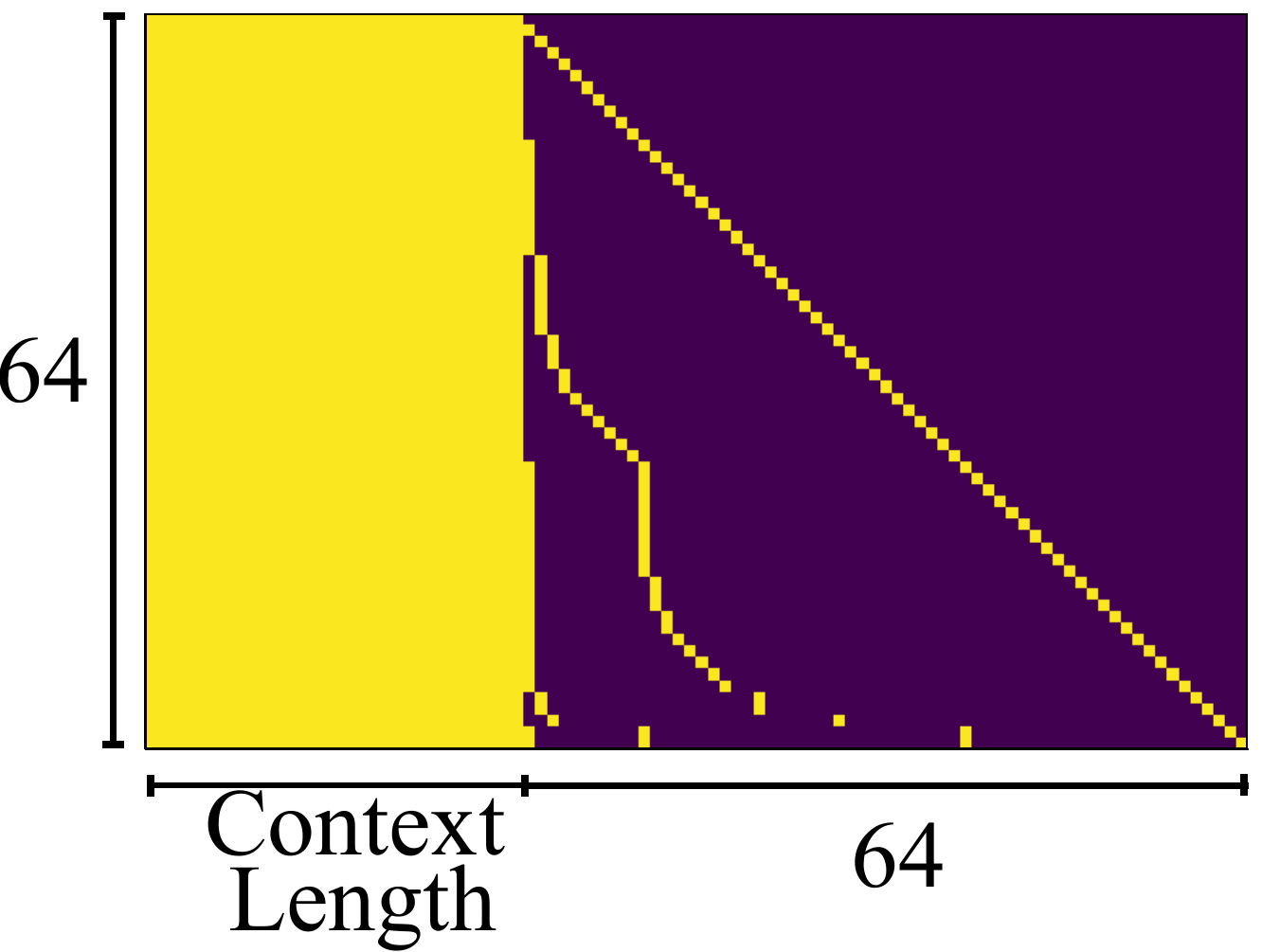}
        \label{fig:medusa_mask}
    }
    
    \subfigure[Draft model~\cite{specinfer}.]{
        \includegraphics[width=0.18\textwidth]{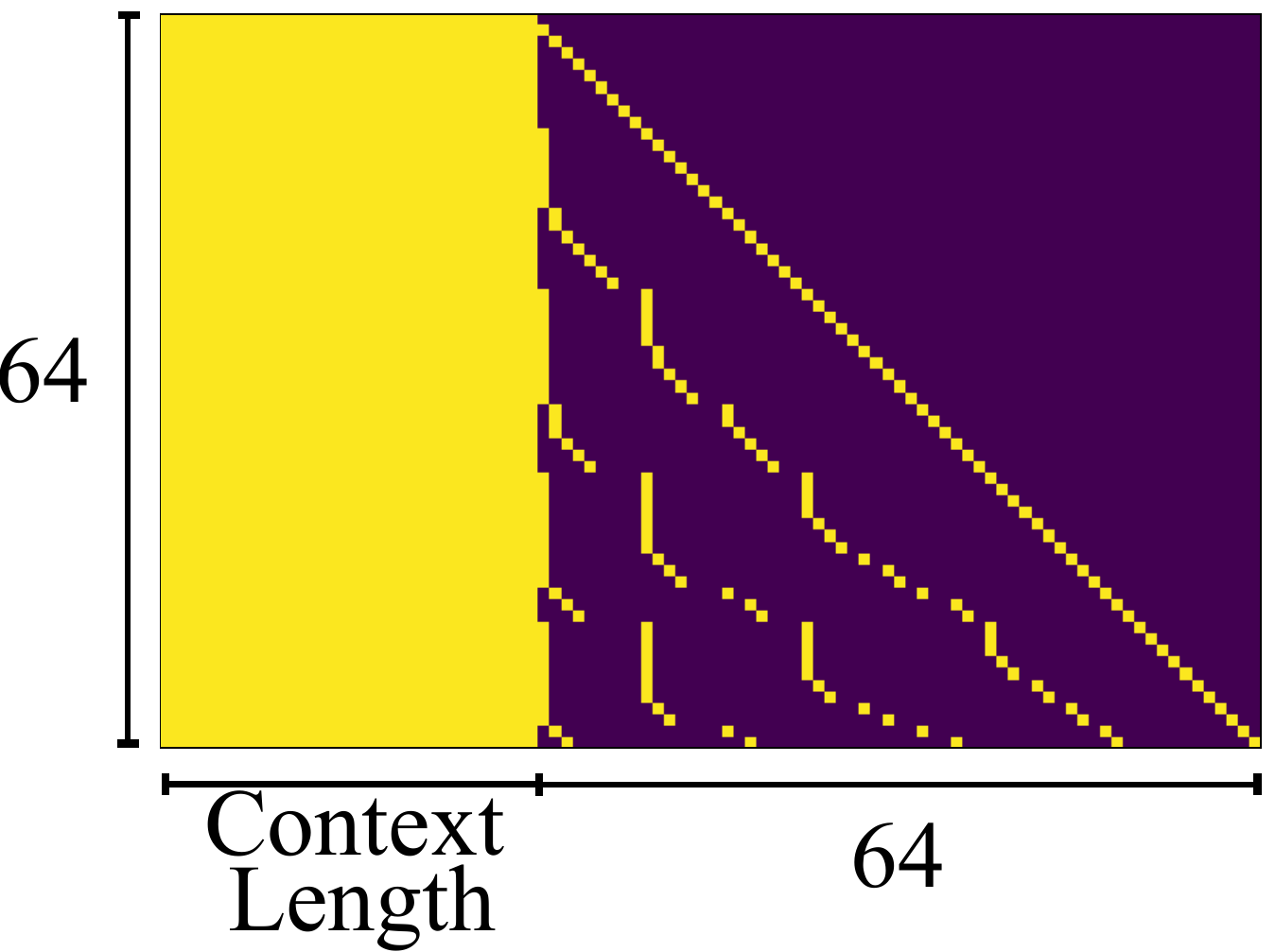}
        \label{fig:draft_model_mask}
    }
    \subfigure[Lookahead~\cite{lookahead_decoding}.]{
        \includegraphics[width=0.18\textwidth]{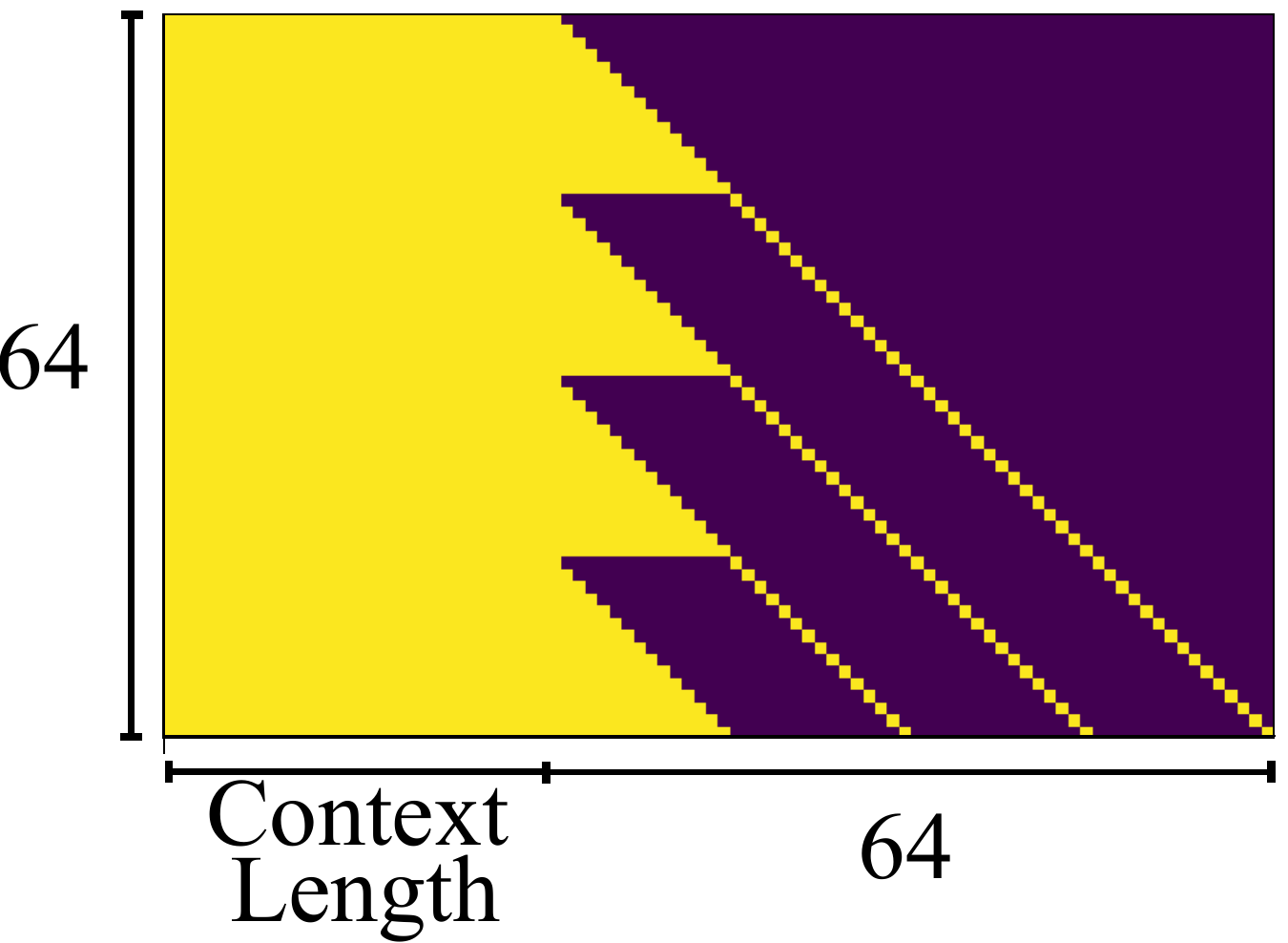}
        \label{fig:Lookahead_mask}
    }
    \vspace{-10pt}
    \caption{Sparsity visualization of the $X = Q \times K^T$ for speculative decoding approaches. Light yellow indicates points that require computation, while dark blue represents points that do not need computation.}
    % \vspace{-15pt}
    \label{fig:medusa_masks}
\end{figure}

To overcome the sequential limitation, speculative decoding is introduced as an alternative. 
To achieve higher parallelism in decoding steps, it primarily employs the Predict-then-Verify mechanism, where predictions for multiple steps are made in parallel and then fallback to the longest validated prefix LLM inference.
There are independent drafting~\cite{specinfer, zhou2023distillspec} and self-drafting~\cite{medusa, eagle,lookahead_decoding} approaches.
The independent drafting employs a smaller model with significantly reduced computation to generate predictions.
The self-drafting make predictions through appending more heads or using Jacobi iteration.
After drafting multiple candidates, these candidate tokens are combined together under a given pattern and put into the 
target model for verification.
Notably, these self-drafting approaches combine the prediction and verification into a single step.
For example, Medusa~\cite{medusa} predicts candidates for future token slots by appending additional heads at the tail of LLMs.
In Figure~\ref{fig:mask_tree}, Head1 has two candidates and Head2 has three candidates.
We aim to verify which combinations between these tokens can be accepted.
Here the verification width in Figure~\ref{fig:mask_tree} is 8 and only a subset of token correlations needs to be calculated in the attention module.
% Determining the verification tree based on a specific verification width is crucial for increasing the acceptance length.

Basically, we observe significant sparsity within the attention module when performing these speculative decoding approaches.
Under a given pattern, a token needs to compute relationships only with certain tokens.
This sparsity is typically handled as dense computation using the mask mechanism.
As illustrated in Figure~\ref{fig:medusa_mask} \ref{fig:draft_model_mask} \ref{fig:Lookahead_mask}, here we present the sparsity visualization of $Q\times K$ result matrix in three popular approaches, namely draft model~\cite{specinfer}, Medusa~\cite{medusa}, and LookAhead~\cite{lookahead_decoding}.
Only the light yellow points represent computations worth performing, highlighting the significant sparsity in the right part.
Softmax and $A\times V$ operation exhibits the same sparsity pattern.

\subsection{Unified Memory Architecture}
\begin{figure}[!t]
    \centering 
    \includegraphics[width=0.48\textwidth]{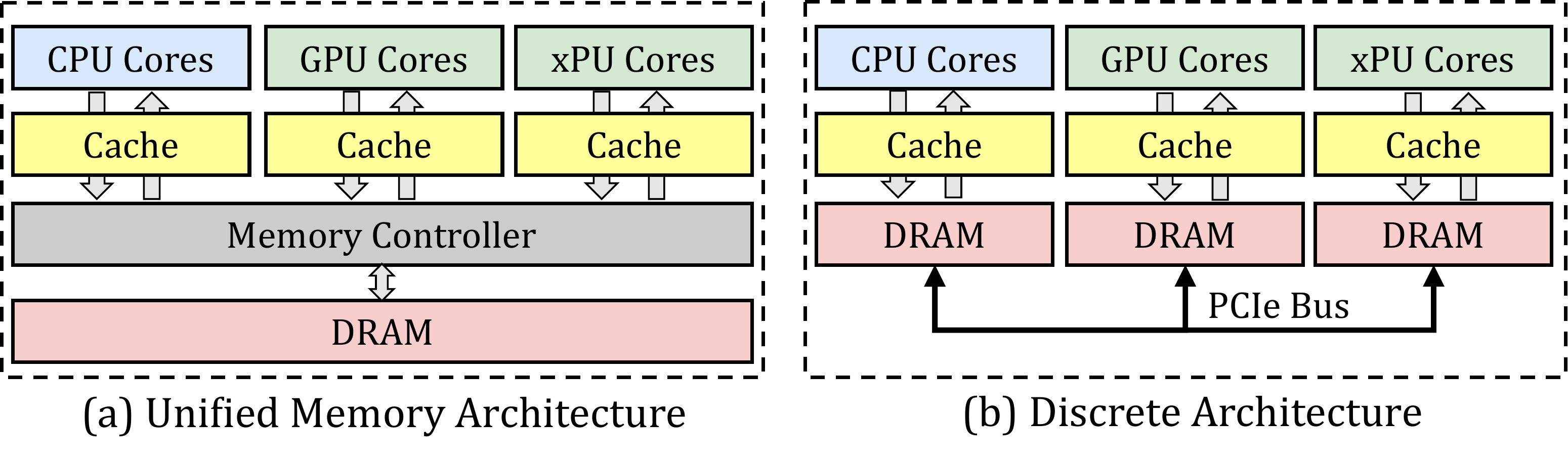}
    \vspace{-10pt}
    \caption{The unified memory architecture and the discrete architecture.}
    % \vspace{-10pt}
    \label{fig:uma}
\end{figure}

End-user devices~\cite{appleA17, appleM4, intelcoreultra, nvjetson, amd, qualcomm}, such as consumer electronics and embedded systems, prioritize power efficiency, cost-effectiveness, and compact design, usually adopting the unified memory architecture.
We show the unified memory architecture and discrete architecture in Figure~\ref{fig:uma}.
End-user devices typically adopt a unified DRAM architecture shared across multiple processing units, rather than discrete memory modules.
This design eliminates explicit data transfer via PCIe bus and enables a zero-copy mechanism between processing units.
While this significantly reduces data synchronization overhead, concurrent memory access still requires manual consistency management by the programmer.
On the NVIDIA Jetson Xavier NX, memory page synchronization between processing units takes less than 0.1 milliseconds when accessing memory written by another unit.
\section{Ghidorah Design}

\subsection{Overview}

\begin{figure}[!t]
    \centering 
    \includegraphics[width=0.42\textwidth]{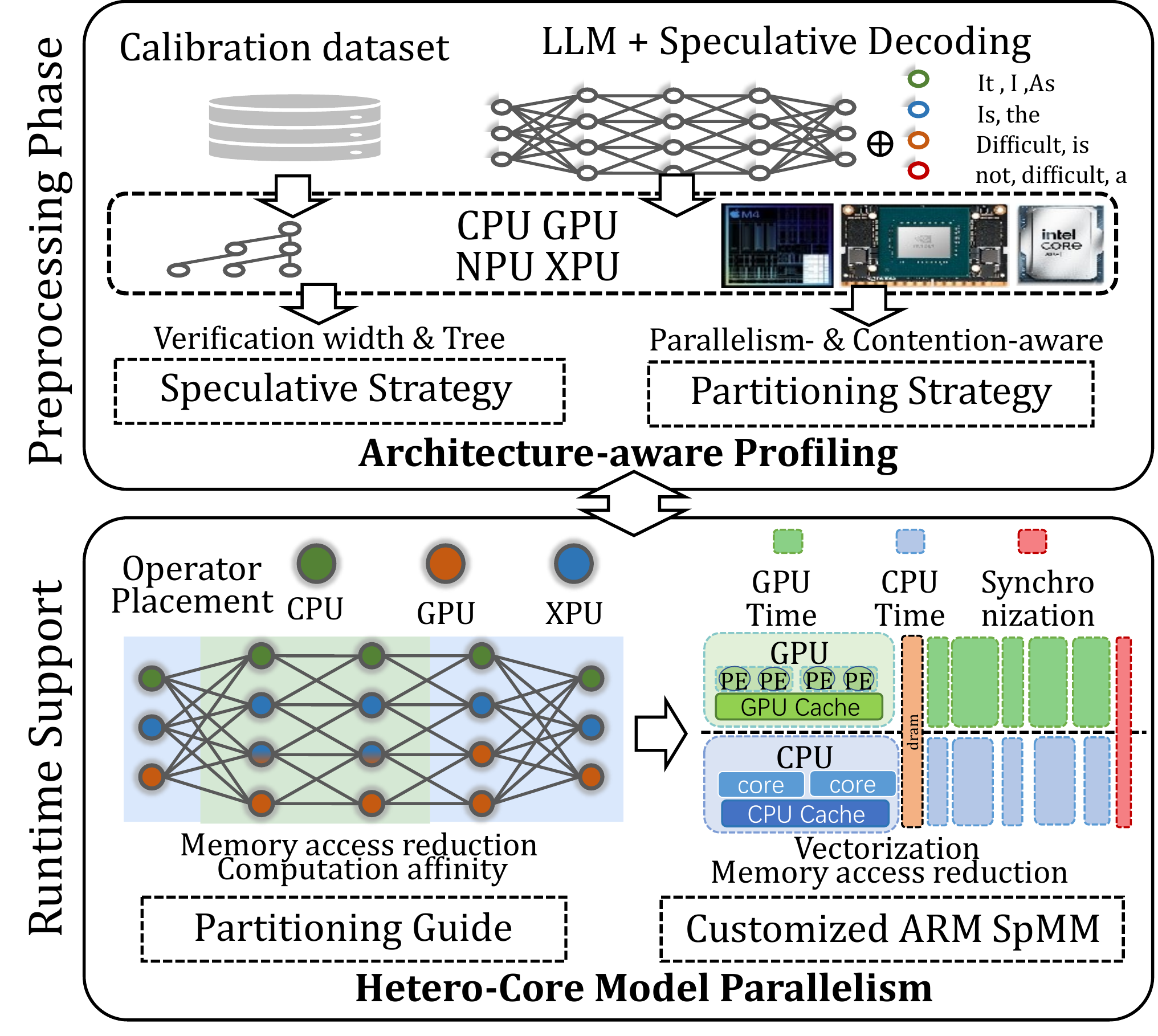}
    \vspace{-10pt}
    \caption{The overview of Ghidorah.}
    \vspace{-5pt}
    \label{fig:overview}
\end{figure}

Ghidorah is specifically designed for single-sample LLM generative task inference on end-user devices with a unified memory architecture.
It primarily focuses on accelerating the dominant decoding phase.
Since the decoding phase is constraint by the memory bandwidth, Ghidorah explores speculative decoding approaches to enhance parallelism and alleviate the memory bandwidth bottleneck.
This manuscript takes the multi-head drafting approach, Medusa~\cite{medusa}, as the default.
Ghidorah distributes the speculative decoding workloads across all processing units and choose the optimal verification length in the end-user device to maximize the inference speed.

Figure~\ref{fig:overview} illustrates the overview of Ghidorah.
Leveraging the unified memory architecture of end-user devices and the computation pattern of speculative decoding, Ghidorah employs a novel hetero-core model parallelism (HCMP) architecture and the architecture-aware profiling (ARCA) approach.
Functionally, the HCMP architecture focuses on providing runtime support, while the ARCA approach is primarily designed for the preprocessing phase.
The HCMP architecture provides guidance for model partitioning and includes customized sparse matrix multiplication optimizations.
Based on the runtime support, the ARCA approach provides the preprocessing phase that runs once before deployment.
It performs an inference process using calibration data as input on multiple heterogeneous processing units in the end-user device.
By comprehensively considering acceptance length, hardware parallelism, and resource contention, this phase formulates speculative and partitioning strategies to maximize performance.

\subsection{Hetero-core Model Parallelism Architecture}
\begin{figure}[!t]
    \centering 
    \includegraphics[width=0.5\textwidth]{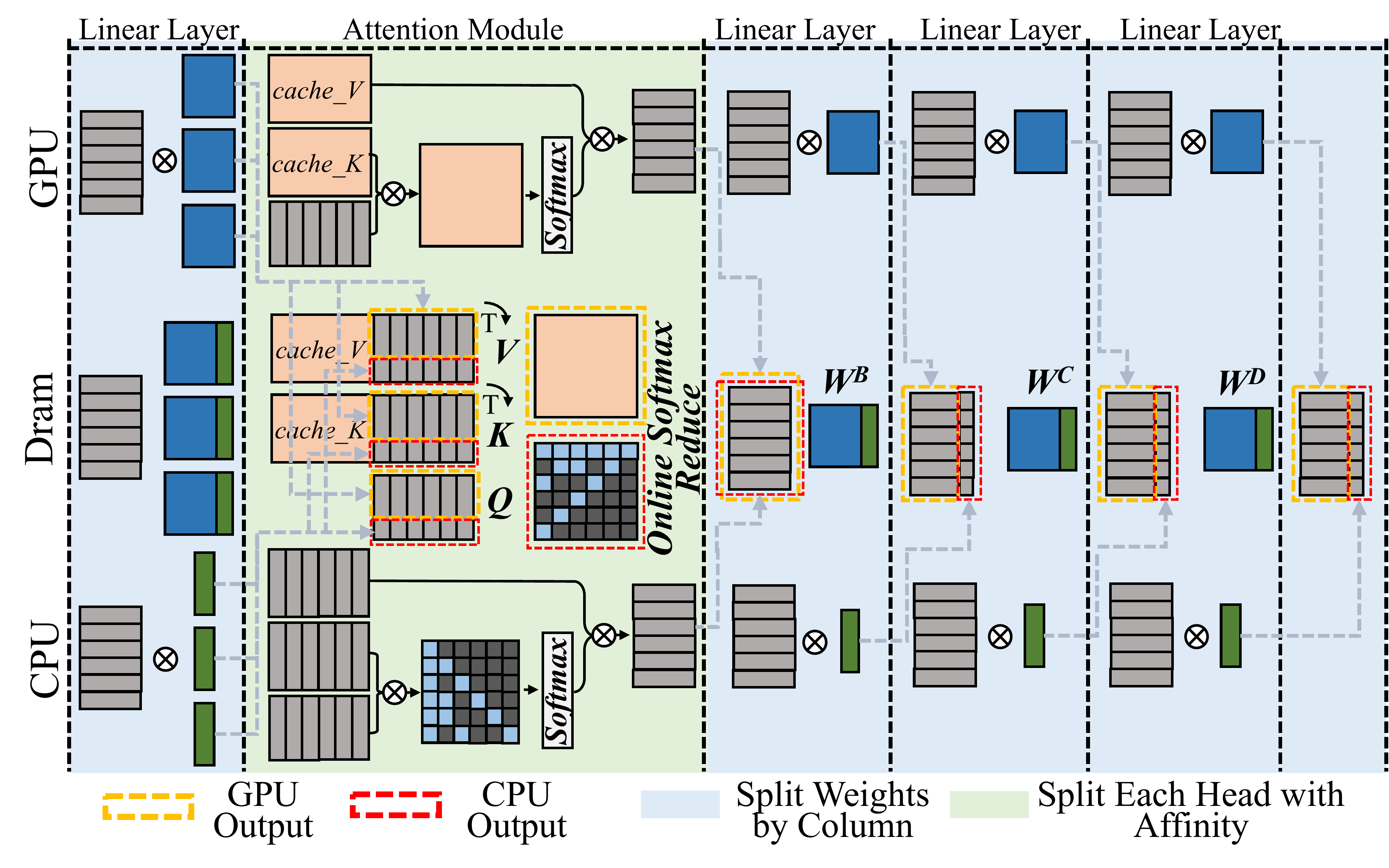}
    \vspace{-10pt}
    \caption{The illustration of Hetero-Core Model Parallelism.}
    \vspace{-5pt}
    \label{fig:hcmp}
\end{figure}

The HCMP architecture is designed for efficiently distributing speculative decoding workloads across multiple heterogeneous processing units organized in the unified memory architecture.
Figure~\ref{fig:hcmp} illustrates the HCMP architecture by presenting an inference example conducted across two processing units, CPU and GPU. 
It shows the memory layout and computation processes separately.
Instead of focusing on reducing inter-device communication, the HCMP architecture prioritizes minimizing memory access and exposing computation affinity across processing units.

\subsubsection{Linear Layer - Split weights by column}
The existing tensor parallelism approach~\cite{shoeybi2019megatron} is designed to distribute workloads across multiple separate devices, requiring explicit inter-device communication to resolve data dependencies. 
Reducing the frequency of tensor synchronization can significantly improve parallel efficiency.
Migrating this setup to end-user devices, for every two linear layers, the weights of the first layer are split by columns and those of the second layer by rows, thus requiring only a single all-reduce synchronization for the two layers.
However, this all-reduce operation necessitates additional memory access.
It must read activations generated by different processing units and write the combined results back.
The HCMP architecture opts to split the weights of all linear layers by columns.
As shown in Figure~\ref{fig:hcmp}, the CPU and GPU receive the same input, multiply it by different columns of weights, and then write the outputs to their designated memory regions without enforcing consistency between them.
This allows the result to be used in the next operation without any additional data access.

\subsubsection{Attention Module - Split each head with affinity}
The HCMP architecture is designed to exploit computation affinity across different processing units, taking into account the hybrid computational intensity inherent in speculative decoding.
The existing tensor parallelism approach divides the computation of the attention module by heads, with each device responsible for processing different heads.
However, since only the correlation between partial tokens requires to be verified in speculative decoding, there exists dense and sparse computation in each head of the attention module.
As shown in Figure~\ref{fig:hcmp}, each attention head can be divided into a dense part — primarily the multiplication with the KV cache — and a sparse part, which mainly involves multiplication with the newly generated key and value.
The dense part will be prioritized for processing units with high parallelism, such as GPU, while the sparse part will be prioritized for processing units with low parallelism, such as CPU.
To achieve load balance, each partition may optionally include a portion of the other part’s computation.
As shown in Figure~\ref{fig:medusa_masks}(b)(c)(d), the left boundary of the sparse part tends to be denser and can be preferentially included in the dense part.

Furthermore, inspired by Ring Attention~\cite{liu2023ring} and FlashAttention~\cite{dao2022flashattention}, we introduce the online softmax technique to extend the span that can be computed continuously.
As shown in Equation~\ref{eq:qkv}, the computation of $O=AV$ on one processing unit cannot begin until $X = QK^T$ on another processing unit is completed, as the softmax operation requires data from all processing units.
With the online softmax technique~\cite{liu2023ring, dao2022flashattention}, each processing unit can have its own softmax operation, aligning their results by applying a scaling factor at the end of the attention module.
This scaling operation can be fused with the reduce operation, introducing almost no overhead.

\subsubsection{Customized ARM SpMM Optimization}

\begin{figure}[!t]
    \centering 
    \includegraphics[width=0.42\textwidth]{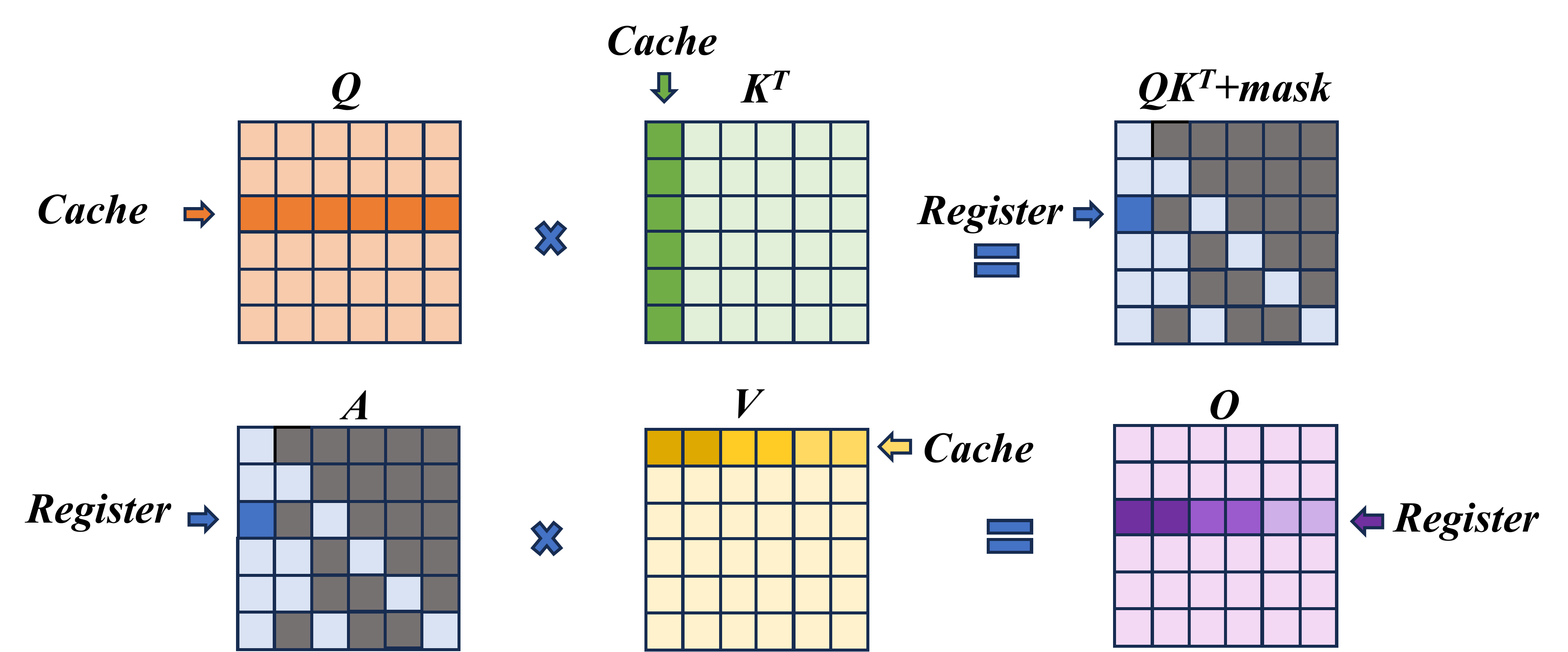}
    \vspace{-10pt}
    \caption{ARM SpMM Optimization.}
    % \vspace{-15pt}
    \label{fig:spmm}
\end{figure}

ARM CPUs are widely used in end-user devices due to their power efficiency feature, and we implement customized computation optimizations for sparse computations in the self-attention module, i.e. $QK^T$ and $AV$.
In details, knowing the token correlations to be verified, we follow the COO sparsity data format to generate the index before performing the inference.
We optimize the sparse operation from two aspects, namely vectorization and memory access reduction.

Figure~\ref{fig:spmm} illustrates the optimization.
In the $QK^T$ computation, the data access for both $Q$ and $K$ is in a row-wise way, thereby supporting continuous data access.
Leveraging NEON instruction set of ARM, we enable the 128-bit long vector instructions to combine memory access in continuous FMA calculations.
For its output matrix, each result value is stored in the register until the final result is fully accumulated, so as to reducing the load/store operations.
In the $AV$ computation, we adjust the execution order to ensure continuous memory access to matrix $V$.
Instead of multiplying with each column of matrix $V$, each non-zero element in the matrix $A[i, j]$ performs multiplication with row j of the matrix $V$ and then accumulates the result to the row i of the output matrix $O$ until the non-zero element in the same row finishes computing and finally gets row i of the output matrix.
Also, since the capacity of registers is limited, we divide this process into blocks, only process partial elements from $V$ to $O$, to enable storing elements in registers and reduce load/store operations.

\subsection{Architecture-aware Profiling Approach}

The ARCA approach comprehensively considers the acceptance length, hardware parallelism and resource contention to finally generate the speculative strategy and the partitioning strategy.
Initially, for a specific speculative decoding approach, we explore verification trees of all verification widths for the optimal acceptance lengths by combining accuracy-based estimation with brute-force search.
Next, we determine the verification width and partitioning ratio for the final deployment, which is both parallelism- and contention-aware.

\subsubsection{Speculative Strategy Determination}

The speculative strategy in Ghidorah consists of both the verification width and the verification tree.
The verification width is the computing budget in a single decoding step, and it is defined as the total number of tokens to be verified, while the verification tree determines the specific combinations of tokens to be verified.
Both the verification width and verification tree influence the acceptance length.
As the verification width increases, it becomes more likely to identify acceptable tokens; however, if the width is too large, the benefits for improving acceptance length becomes limited.
Also, some specific verification routes are more likely to be accepted. For instance, the closer the head is to HEAD0, the easier its tokens are accepted.
We design verification trees for different verification widths.

\begin{figure}[!t]
    \centering 
    \includegraphics[width=0.48\textwidth]{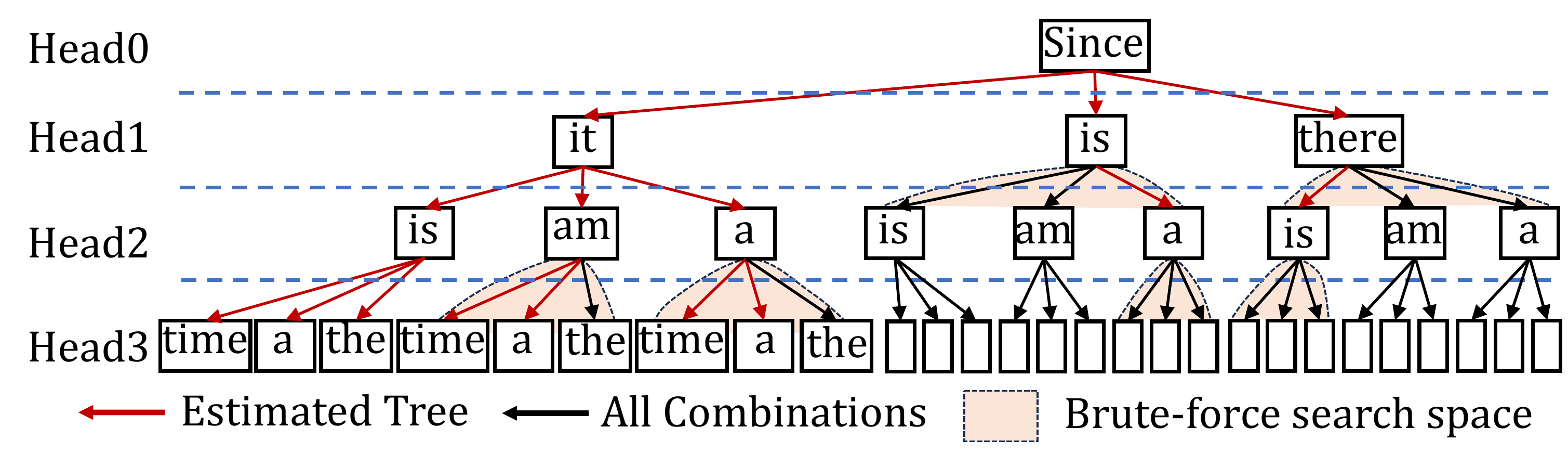}
    \vspace{-10pt}
    \caption{The verification tree determination with a verification width of 16.}
    % \vspace{-15pt}
    \label{fig:tree_design}
\end{figure}

Figure~\ref{fig:tree_design} illustrates the tree design process with a verification width of 16 (nodes), involving candidate tokens generated from 4 heads of Medusa~\cite{medusa}.
In details, we use a calibration dataset and generate the accuracies of the top predictions of different heads.
Then, we estimate the accuracy of a candidate sequence by multiplying the accuracies of its related tokens.
Then we can get the overall expected acceptance length by accumulating the expected acceptance length of different candidate sequences.
Maximizing the acceptance length expectation, we can add nodes with the highest accuracies one by one until reaching the given verification length.
As in Figure~\ref{fig:tree_design}, the estimated tree is indicated by red lines.
We further employ the brute-force search based on the estimated tree and compare their real acceptance lengths to determine the final tree.
In Figure~\ref{fig:tree_design}, we search leaf nodes and nodes in the same level.

\subsubsection{Parallelism-Aware Profiling}
% 感觉其实是8，16，24，32，40，48，56，64 比较好。
The ARCA approach is parallelism-aware.
Modern processing units leverage vectorization techniques to boost computational efficiency, leading to stepwise performance improvements in LLM inference latency as batch size increases, a phenomenon referred to as wave quantization~\cite{wavequant} on NVIDIA hardware. Consequently, different processing units exhibit distinct performance sweet spots.
In our preliminary experiments, we observe that setting candidate verification widths to powers of two, specifically 2, 4, 8, 16, 32, and 64, aligns well with the vectorization capabilities of these devices, resulting in more efficient execution. This observation can be explained as follows.
The execution time of general matrix multiplication (GeMM) operations, including linear and attention, generally dominates LLM workloads.
When applying model parallelism, the hidden state dimension is partitioned across processing units, while the token dimension (verification length) remains unchanged.
As a result, the split GeMM operations executed on different processing units share the same token dimension and collectively span the hidden state dimension.
Next, we determine their verification trees and check these speculative strategies with the runtime support.
This determination process is contention-aware.

\subsubsection{Contention-Aware Profiling}
Although the partitioning guidance of the HCMP architecture will not increase the overall memory access within a single step of speculative decoding, the simultaneous access from different processing units will lead to resource contention problem.
Since this contention affects different processing units at varying levels, it is difficult to determine the partitioning ratio.
In this way, the ARCA approach initializes the partitioning strategy based on the individual execution times of different processing units and determines the final partitioning strategy for a given verification width through gradual adjustments.
Besides, as the context (KV cache) length will impact the ratio of sparsity in the attention module, we additionally adjust the context length for the attention module to support its dynamic partitioning.
\section{Evaluation}

We implement a prototype system of Ghidorah in C/C++.
We utilize the CUDA kernels from the highly optimized Transformer inference system FasterTransformer~\cite{nvidia2023fast} for the GPU and rebuild the ARM CPU code based on CTranslated2~\cite{CTranslate2}.
In this section, we evaluate Ghidorah from two aspects.
1) \textbf{Verification Efficiency:} we evaluate the acceptance length of our speculative strategies, and 
2) \textbf{Overall Performance:} we compare the single-sample inference of Ghidorah with that of state-of-the-art approaches. 
Here we only focus on the dominant decoding phase of the LLM inference.

\subsection{Experimental Setup}

\textbf{Model and Speculative Decoding.}
We evaluate Ghidorah with Vicuna-7B~\cite{MT-bench} and speculative decoding approach Medusa~\cite{medusa}.
Medusa originally offers a 5-head version of Vicuna-7B and we design our own verification trees to evaluate the acceptance length under different verification widths.
Vicuna-7B is a fine-tuned version from LLaMA-7B~\cite{touvron2023llama}, which is a widely-recognized model architecture in the AI community.

\textbf{Datasets.}
We evaluate Ghidorah for the acceptance length across a broad spectrum of datasets, including: 1) MT-Bench~\cite{MT-bench}: a diverse collection of multi-turn questions, 2) GSM8K~\cite{GSM8k}: a set of mathematical problems, 3) HumanEval~\cite{HumanEval}: code completion and infilling tasks, and 4) MBPP~\cite{MBPP}: instruction-based code generation.

\textbf{Node testbed.}
We use NVIDIA Jetson NX, which is a CPU-GPU edge device with unified memory architecture.
It consists of an embedded 384-core Volta GPU and a 6-core ARM v8.2 CPU.
We lock the GPU at 204MHz and the CPU at 1.9GHz to simulate end-user devices with more balanced capabilities of heterogeneous processing units.
Specifically, we use g++ 7.5, cuda 10.2, arm performance library 24.10.
% The memory is a 16GB LPDDR4x with a bandwidth of 59.7 Gbps. 

\textbf{Metrics.}
We use the acceptance length and decoding throughput as the evaluation metrics.
The acceptance length refers to the average number of tokens that the target model accepts in a single decoding step.
The decoding throughput is the number of tokens generated in a fixed interval.
% and this comes from both speculative decoding and parallel computing.
% Due to memory constraints, we execute only one decoding layer, preventing us from obtaining the complete speedup ratio. 
% Thus, the speedup ratio utilized in this paper is based on the acceptance length and the ratio of differing decoding times.

\textbf{Baseline Methods.}
We compare Ghidorah with approaches:
\begin{itemize}
    \item \textbf{Sequential:} The sequential decoding approach running on the GPU, which is generally limited by the algorithmic parallelism.
    \item \textbf{Medusa:} The Medusa decoding approach running on the GPU and it adopts our verification trees in given verification widths.
    \item \textbf{Medusa+EM (Medusa + \underline{E}dgeNN~\cite{edgenn} + \underline{M}egatron)~\cite{shoeybi2019megatron}:} 
    The Medusa decoding approach is distributed across the CPU and GPU using the partitioning guidance of the tensor parallelism from Megatron-LM. Additionally, we apply zero-copy optimization to reduce the synchronization overhead and determines the partitioning ratio based on the execution times of the processing units as in EdgeNN.
\end{itemize}

% \vspace{-10pt}
\subsection{Acceptance Length Evaluation}
\begin{table}[!t]
\caption{Acceptance length under given verification widths.}
\label{tab:acc-length}
\centering
\resizebox{0.9\linewidth}{!}{
\begin{tabular}{c|c|c|c|c|c|c|c|c}
\hline
\multicolumn{2}{c|}{Verification Width} & 1 & 2 & 4 & 8 & 16 & 32 & 64 \\ \hline
\multirow{4}{*}{Dataset} 
& MT-bench & 1 & 1.72 & 2.28 & 2.59 & 2.93 & 3.19 & 3.34 \\ \cline{2-9}
& GSM8K & 1 & 1.76 & 2.43 & 2.69 & 3.08 & 3.34 & 3.56 \\ \cline{2-9}
& MBPP & 1 & 1.78 & 2.54 & 2.89 & 3.27 & 3.55 & 3.74 \\ \cline{2-9}
& Human-eval & 1 & 1.77 & 2.49 & 2.8  & 3.19 & 3.48 & 3.71 \\ \hline
\end{tabular}
}
\vspace{-10pt}
\end{table}

We present the acceptance lengths for various datasets under different verification widths.
Notably, MT-bench is used as the calibration dataset for determining verification trees.
Subsequently, these verification trees are applied to three other datasets to generate their acceptance lengths.
Table~\ref{tab:acc-length} shows the acceptance length results.
Larger verification widths generally result in longer acceptance lengths, while the benefit becomes weak as verification widths grow and come at the cost of significantly increased computational effort.
Furthermore, we find that these verification trees exhibit strong generality, achieving even better acceptance lengths on GSM8K, MBPP, and Huamn-eval, when migrated from MT-Bench.
This can be attributed to MT-Bench's comprehensive nature, as it includes conversations across eight different types.

\begin{figure*}[!t]
    \centering 
    \includegraphics[width=0.95\textwidth]{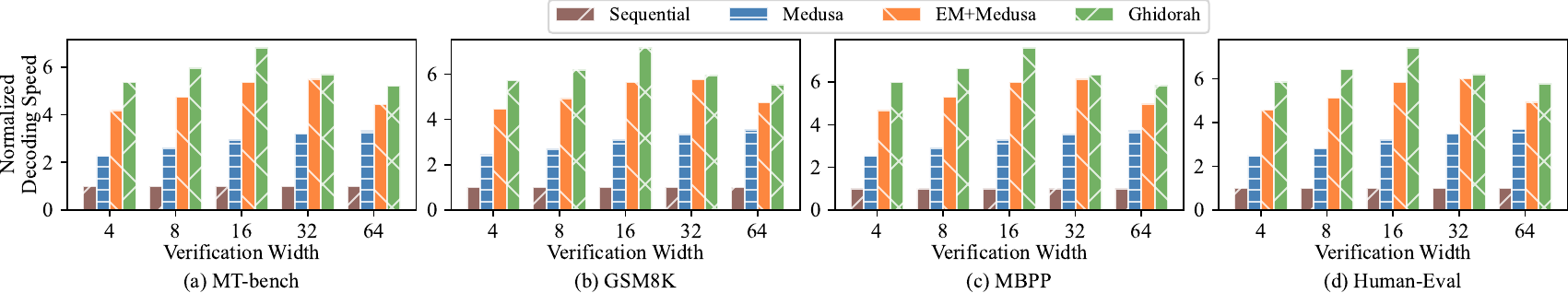}
    \vspace{-10pt}
    \caption{The overall performance under different verification widths.}
    \vspace{-5pt}
    \label{fig:general_perf}
\end{figure*}

\subsection{Overall Performance Evaluation}

% 和sequential对比：[5.98227378 6.64942956 7.60057791 6.33372212 5.83276936]
% 和Medusa对比：[2.34112821 2.28911323 2.3141255 1.77972337 1.56229027]
% 和Medusa + EM 对比： [1.28250923 1.25435467 1.27043089 1.0328675 1.17139669]

This section compares the decoding throughput of Ghidorah against that of baseline methods with different verification widths from 4 to 64.
We select a context length (KV cache length) of approximately 256.
Figure~\ref{fig:general_perf} shows the normalized results.
In all cases shown in Figure~\ref{fig:general_perf}, Ghidorah presents significant speedup in the decoding throughput.
Especially when the verification width is 16, Ghidorah's speedup can reach up to 7.6$\times$ compared with the sequential decoding.

The impressive performance improvement of Ghidorah can be attributed to two main factors:
First, the increased acceptance length enabled by speculative decoding.
Since the sequential decoding approach is memory-bandwidth-bound, its execution time is similar to that of Medusa, and the acceptance length is fixed at 1.
However, the Medusa method generates multiple tokens in a single step and requires the similar execution time; specifically, as the verification width increases, the acceptance length of Medusa also grows.
Second, the collaborative operation of multiple heterogeneous processing units can better conduct the Medusa decoding workload.
When comparing Medusa, Medusa+EM, and Ghidorah, we find that utilizing heterogeneous processing unit yields a better acceleration effect than merely running Medusa on the GPU.
Taking MBPP dataset as the instance, Ghidorah achieves an average speedup of 2.06$\times$ compared to running Medusa solely on the GPU, and an average speedup of 1.20$\times$ compared to Medusa+EM.
The improvement over Medusa+EM is attributed to reduced memory access and improved computing affinity.

Longer acceptance lengths do not always result in better throughput, making it essential for the ARCA approach to balance the algorithmic parallelism with hardware capabilities.
As shown in Figure~\ref{fig:general_perf}, Medusa and Ghidorah achieve their best throughput results at verification width of 64 and 16 respectively.
This is because CPU and GPU have different sweet spots.
The GPU maintains a similar execution time from 4 to 64 verification width and achieves a longer acceptance length at verification width of 64, while the CPU can only maintain a similar execution time from 4 to 16 verification width.
Since increasing verification width from 16 to 64 results in only 0.47 increase in the acceptance length, Ghidorah achieves the optimal throughput at a verification width of 16.

\subsection{Dynamic Partitioning and Sparse Optimization}

\begin{figure}[!t]
    \centering
    \subfigure[The attention module performance.]{
        \includegraphics[width=0.32\textwidth]{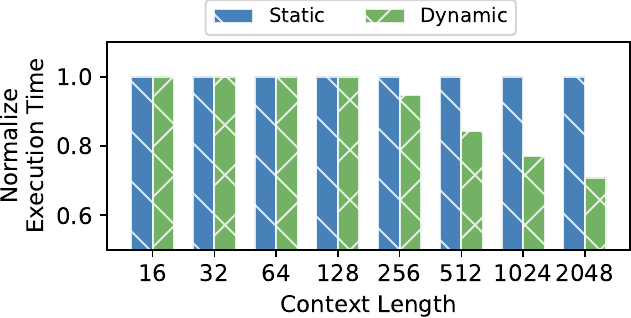}
        \label{fig:attn_perf}
    }
    \hspace{5mm}
    \subfigure[ARM Optim.]{
        \includegraphics[width=0.095\textwidth]{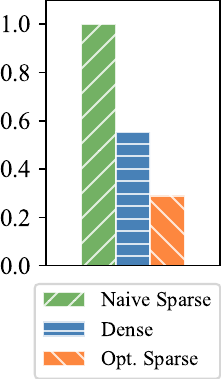}
        \label{fig:spmm_res}
    }
    \vspace{-5pt}
    \caption{Performance of Sparse Optimization}
    \vspace{-10pt}
    \label{fig:sparse}
\end{figure}

% Dynamic/Static [1, 1, 1, 1, 0.94598618, 0.8432132, 0.77075987, 0.70815065]

% Dense / Naive Sparse 0.55168757
% Opt. Sparse / Naive Sparse 0.28993387
% Opt. Sparse / Dense 0.52553997
% Since these two techniques are primarily associated with the attention module, we individually present its performance with varying context lengths.

This section presents the effectiveness of the dynamic partitioning and the sparsity optimization.
Since the dynamic partitioning merely impact the attention module, Figure~\ref{fig:attn_perf} represents the attention module performance with a verification length of 64.
Notably, the attention module becomes dominant in the Transformer model as the context length increases.
Static refers to that all sparse computations are processed on the CPU and all dense computations are processed on the GPU.
Dynamic indicates that sparse computations are primarily handled by the CPU, while dense computations are primarily handled by the GPU, with dynamic partitioning guided by profiling results.
Dynamic partitioning demonstrates obvious improvements when dealing with large context lengths.

Figure~\ref{fig:spmm_res} illustrates the performance of the sparse component executed using three different strategies: a naive sparse implementation, our optimized sparse implementation, and the dense implementation that treats the sparse computation as dense.
The execution time of optimized sparse implementation exceeds naive sparse and dense implementations by 3.49$\times$ and 1.90$\times$ respectively.
Also, if we do not optimize the sparse computation, the naive sparse execution cannot outperform the dense one.
\section{Related Work}

% \noindent\textbf{Speculative Decoding.} 
% Speculative Decoding represents an innovative paradigm designed to accelerate LLM inference. 
% The three primary speculative decoding approaches differ fundamentally in their generation of prediction sequences. 
% Specifically, Specinfer~\cite{specinfer} and Glide~\cite{glide} employ draft models for sequence prediction, while Medusa\cite{medusa} and Eagle~\cite{eagle} utilize additional feed-forward networks (FFNs). 
% In contrast, Lookahead~\cite{lookahead_decoding} and Rest~\cite{rest} leverage contextual information and databases for sequence generation. 
% Although our method has been primarily validated using Medusa, it is also suitable for other speculative decoding approaches.

% 是不是改成投机采样的相关优化工作比较好？
% SpecInfer~\cite{specinfer}
% Glide~\cite{glide}
% LLMcad ~\cite{llmcad}
% SpecExec ~\cite{specexec}
% BASS ~\cite{Bass}

% \noindent\textbf{System Assisted by Speculative Decoding.} 
% Recent research has focused on optimizing speculative decoding. 
% Sequoia introduced adaptive tree structures that consider both device memory constraints and model size~\cite{sequoia}. 
% SpecInfer and Glide focused on modifying the draft model inference process to improve compatibility with the target model~\cite{specInfer, glide}. 
% For server-side optimization, BASS developed an enhanced attention kernel specifically for batched speculative decoding requests~\cite{Bass}. 
% However, these studies have not adequately addressed optimization strategies for unified memory systems in end-user devices.

\noindent\textbf{Collaborative Execution of Transformer.}
Tensor Parallelism~\cite{kwon2023efficient} and Pipeline Parallelism~\cite{agrawal2024taming} effectively addresses memory limitations by accumulating memory from more devices, gaining widespread adoption in inference tasks within datacenters.
They are also used in end-user environments.
Galaxy~\cite{ye2024galaxy} hides communication into computation in the tensor parallelism approach for edge devices, while DeTransformer~\cite{wei2024communication} fundamentally mitigates data dependency by decoupling the original model.
Petals~\cite{borzunov2022petals} utilizes pipeline parallelism to distribute LLM inference across geographically distributed end-user devices.

% Data Parallelism~\cite{zero} remains the predominant distributed training approach in datacenters. 
% While Pipeline Parallelism~\cite{huang2019gpipe} was later introduced to address memory constraints in training large-scale transformer-based models , pipeline bubbles hinder it. 

\noindent\textbf{Integrated CPU-GPU Scheduling and Execution.} 
There has been a large amount of prior work focusing on improving the device utilization on integrated CPU-GPU architecture.
CoDL~\cite{codl} and EdgeNN~\cite{edgenn} perform collaborative convolutional neural network inference on mobile CPU and GPU with unified memory architecture.
Asymo~\cite{AsyMo} schedules workloads based on the computing power of asymmetric mobile CPU cores to achieve higher throughput.
Pipe-It~\cite{wang2019highhigh} achieves higher throughput by organizing ARM big.Little cores in a pipelined manner.
However, none of these approaches are designed for Transformer-based models.

\noindent\textbf{Speculative Decoding System.} 
Speculative Decoding represents an innovative paradigm designed to accelerate LLM inference. 
Existing LLM inference systems~\cite{vllm} start to integrate speculative decoding approaches and they treat them as ordinary LLM inference with a large batch size.
Sequoia~\cite{sequoia} explores ultra-wide verification width for advanced hardware in cloud. 
LLMCad~\cite{llmcad} adopts the independent drafting speculative decoding approach to accommodate memory-constrained devices. 
It utilizes a smaller model to generate drafts, thereby reducing data loading time for the larger model.

\section{Conclusion}

Ghidorah focuses on deploying LLM inference on end-user devices with the unified memory architecture.
To fully exploit the hardware capability, Ghidorah leverages speculative decoding approaches to enhance the algorithmic parallelism and then distributes this novel workload across multiple heterogeneous processing units.
With the help of the HCMP architecture and ARCA approach, it can achieve up to 7.6$\times$ speedup in the dominant decoding phase.
In the future, Ghidorah will be open-sourced and extended to support more speculative decoding approaches and a wider range of end-user devices.

% be open-sourced integrate

\bibliographystyle{IEEEtran}
\bibliography{IEEEabrv, reference}

\end{document}